\documentstyle[twoside,fleqn]{article}

\newcommand{\bls}[1]{\renewcommand{\baselinestretch}{#1}}
\def\noi{\noindent}

\makeatletter
\renewcommand{\section}{\@startsection{section}{1}{0pt}%
        {-3.5ex plus -1ex minus -.2ex}{2.3ex plus .2ex}%
        {\large\bf\protect\raggedright}}

\renewcommand{\subsection}{\@startsection{subsection}{2}{0pt}%
        {-3ex plus -1ex minus -.2ex}{1.4ex plus .2ex}%
        {\normalsize\bf\protect\raggedright}}

\renewcommand{\thesubsubsection}%
        {\arabic{section}.\arabic{subsection}.\arabic{subsubsection}.}
\renewcommand{\@oddhead}{\raisebox{0pt}[\headheight][0pt]{%
   \vbox{\hbox to\textwidth{\rightmark \hfil \rm \thepage \strut}\hrule}}}
\renewcommand{\@evenhead}{\raisebox{0pt}[\headheight][0pt]{%
   \vbox{\hbox to\textwidth{\thepage \hfil \leftmark \strut}\hrule}}}
\newcommand{\heads}[2]{\markboth{\protect\small\it #1}{\protect\small\it #2}}
\newcommand{\Acknow}[1]{\subsection*{Acknowledgement} #1}
\makeatother

\newcommand{\Title}[1]{\noi {\Large #1} \\}
\newcommand{\Author}[2]{\noi{\large\bf #1}\\[2ex]\noi{\it #2}\\}
\newcommand{\Abstract}[1]{\vskip 2mm \begin{center}
     \parbox{16.4cm}{\small\noi #1} \end{center}\bigskip}

\newcommand{\sect}[1]{Sec.\,#1}
\def\nq{\hspace{-1em}}
\def\nqq{\hspace{-2em}}
\def\nhq{\hspace{-0.5em}}

\def\cm{\hspace{1cm}}
\def\inch{\hspace{1in}}
\def\Ref#1{Ref.\,\cite{#1}}

\def\eq{Eq.\,}
\def\eqs{Eqs.\,}
\def\beq{\begin{equation}}
\def\eeq{\end{equation}}
\def\bear{\begin{eqnarray}}
\def\al{&\nhq}
\def\lal{&&\nqq {}}               
\def\bearr{\begin{eqnarray} \lal}
\def\ear{\end{eqnarray}}
\def\earn{\nonumber \end{eqnarray}}
\def\dst{\displaystyle}
\def\tst{\textstyle}

\def\nn{\nonumber\\ {}}

\def\nnn{\nonumber\\ \lal }

\def\yy{\\[5pt]}

\def\eql{\al =\al}

\def\e{{\,\rm e}}
\def\d{\partial}
\def\re{\mathop{\rm Re}\nolimits}
\def\im{\mathop{\rm Im}\nolimits}

\def\diag{\mathop{\rm diag}\nolimits}
\def\dim{\mathop{\rm dim}\nolimits}
\def\const{{\rm const}}
\def\Half{{\dst\frac{1}{2}}}
\def\half{{\tst\frac{1}{2}}}

\newcommand{\aver}[1]{\langle \, #1 \, \rangle \mathstrut}

\def\intl{\int\limits}

\catcode`\@=11 \@addtoreset{equation}{section}\catcode`\@=12
\newcommand{\be}{\begin{equation}}
\newcommand{\ee}{\end{equation}}
\newcommand{\ba}{\begin{eqnarray}}
\newcommand{\ea}{\end{eqnarray}}
\newcommand{\p}{\partial}

\newcommand{\R}{{\sf R\hspace*{-0.8ex}\rule{0.1ex}{1.5ex}\hspace*{0.8ex}}}
\newcommand{\N}{{\sf N\hspace*{-1ex}\rule{0.1ex}{1.4ex}\hspace*{1ex}}}

\newcommand{\C}{{\sf C\hspace*{-0.87ex}\rule{0.1ex}{1.4ex}\hspace*{0.87ex}}}

\heads{Vladimir D. Ivashchuk}
{Wick Rotation, Regularization of Propagators
by a Complex Metric and Multidimensional Cosmology}

\bls{1.02}

\begin{document}
\thispagestyle{empty}
\twocolumn[
\noi \unitlength=1mm
\begin{picture}(174,8)
       \put(31,8){\shortstack[c]
       {RUSSIAN GRAVITATIONAL SOCIETY\\
       INSTITUTE OF METROLOGICAL SERVICE \\
       CENTER FOR GRAVITATION AND FUNDAMENTAL METROLOGY }        }
\end{picture}
\begin{flushright}
                                         RGS-VNIIMS-97/04\\
                                         hep-th/9705008 \\
     	        To appear in  {\it Grav. and Cosmol.} {\bf 3}, 1(9), (1997)
\end{flushright}
\medskip

\Title{WICK ROTATION, REGULARIZATION OF PROPAGATORS   \yy
       BY A COMPLEX METRIC AND MULTIDIMENSIONAL COSMOLOGY}

\Author{Vladimir D. Ivashchuk }
{Centre for Gravitation and Fundamental Metrology, VNIIMS,
     3--1 M. Ulyanovoy St., Moscow 117313, Russia}

\Abstract
{The Wick rotation in quantum field theory is considered in terms of
analytical continuation in the signature matrix parameter $w = \eta_{00}$.
Regularization of propagators by a complex metric parameter
in most cases preserves
(i) the convergence of Feynmann integrals (understood
as Lebesgue integrals)  if the corresponding integrals of Euclidean
theory are convergent;
(ii) the regularity of propagators in the coordinate representation
if there is regularity in the Euclidean case.
The well-known covariant regularization
by a complex mass does not in general satisfy these conditions.
Theories with a large family of propagators regularized
by complex metric were previously considered by the author, and
analogues of the Bogoliubov-Parasiuk-Hepp-Zimmermann theorems were proved.
This paper shows that in the case of multidimensional
cosmology describing the evolution of $n$ spaces $M_i$,
$i = 1, \ldots, n$, the Wick rotation in the minisuperspace
may be performed by analytical continuation in the dimensions
$N_i = \dim M_i$ or in the dimension of the time submanifold $M_0$.}

] 

\section{Introduction}

The divergences occuring in quantum field theory (QFT)
and quantum gravity [1--9]
(when certain perturbation schemes are considered) may be divided into two
groups. The first one is related to the fact that series in
coupling constants in QFT usually  diverge (they are asymptotical ones
\cite{GZ}).  (Recently a new perturbation theory with absolutely convergent
series was considered in \cite{BSS}.)
Divergences from the second group are the so-called ``infrared"
and ``ultraviolet" divergences in Feynmann integrals. In
renormalizable theory they may be removed by applying
certain R-operation procedures [1, 4--7]
(with  or without intermediate regularization).
Some of the latter may be called ``pseudo-Euclidean" \cite{Z,I3,I4} since
they occur in space-times with non-Euclidean signatures.
Let us give a simple example. Consider the covariantly
regularized propagator of free massive scalar field in $\R^D$ \cite{Bog}:
\be 
\frac{i}{- p^2_M - m^2 +i \varepsilon}, \qquad \varepsilon > 0,
\ee
where $p^2_M = - p_0^2 + \vec{p}^{\ 2}$, $m^2 > 0$, then the regularization
(1.1) does not guarantee the convergence of Feynmann integrals (understood
as Lebesgue integrals) even if the corresponding integrals of Euclidean
theory are convergent \cite{I3} (see also Subsec.\,2.2.4 below). The
covariant regularization does not guarantee the propagator regularity
in the coordinate representation even if they are regular
in the Euclidean case. For $m^2 > 0$, $\varepsilon > 0$, $D \geq 4$
the Fourier image of (1.1) in  the Schwarz  space ${\cal S}'(\R^D)$
is not a regular distribution \cite{I3} (see also Subsec. 2.2.3 below).

There exists an alternative scheme of propagator regularization
\cite{I1}-\cite{I4} (see also \cite{BK}-\cite{EOR}).
This scheme is free of the above disadvantages and
is based on the complex metric regularization.
In this scheme we deal with a complex metric on $\R^D$
\be  
{\eta_{ab}}(w) dx^a \otimes dx^b = w dt \otimes dt +
\sum_{i=1}^{D-1} dx^i \otimes dx^i,
\ee
$w \neq 0$. For non-negative $w$ the metric (1.2) provides
regularization of (singular) Green functions
(corresponding to the Minkowski space limit $w \to -1 +i0$).
For example, the massive scalar field propagator regularized
by the complex metric (1.2) has the form \cite{I4}
\be  
D(p,w,m) = \frac{w^{-1/2}}{w^{-1} p_0^2 +  \vec{p}^{\ 2} +  m^2}.
\ee
Thus $w$ may be considered as a holomorphicity parameter
(Wick parameter) for $w$-correlators belonging to
${\cal S}'(\R^D)$. It should be noted that recently the
interest to this problem was greatly stimulated
by the paper of J. Greensite \cite{Greens} (see also \cite{CGr}).

The aim of this paper is to overview the author's earlier results
concerning the regularization of propagators by a complex metric
(\cite{I1}-\cite{I4}) and to show a possible application
of this scheme to multidimensional quantum cosmology.

The plan of this note is as follows. \sect 2 considers the complex
metric regularization in QFT. \sect 3 discusses a gravitational model
on the manifold $M_{0}  \times M_{1} \times \ldots \times M_{n}$.
The midisuperspace metric depends on the dimensions $N_0 = \dim M_0$ and
$N_i = \dim M_i$ and may be used to regularize the cosmological
pseudo-Euclidean minisupermetric corresponding to  $N_0 = 1$
in two ways: (i) $N_0$ is complex and $N_i$ are real;
(ii) $N_0 = 1$ and $N_i$ are complex ($i > 0$).

\section{Wick rotation in quantum field theory}

\subsection{General prescription}

We consider the Wick rotation in QFT
in terms of analytical continuation of the signature parameter.
Let
\be  
{\eta_{ab}}(w) = {\rm diag}(w,1,...,1)
\ee
be a diagonal $D \times D$-matrix, $a,b = 0,...,D-1$, $D \geq 2$,
where
\be  
      w \in \C \setminus (- \infty, 0] \equiv \Omega
\ee
is a complex parameter (Wick parameter).

The point  $w =1$  corresponds to Euclidean space (E). As will
be shown below, the limits $w \to -1 \pm i0$  correspond to the
Minkowski spaces $M_{\pm}$ with a ``right" or ``wrong"
direction (arrow) of time, respectively.

(In \cite{I3,I4} we used the notation
\be  
w = - a^{-1} = - \exp(-i \varepsilon).)
\ee

Consider a self-interacting scalar field with the action
\bearr  
{S}(\varphi,w) = \int d^{D} x\, \sqrt{\eta(w)} \{ \frac{1}{2}
\eta^{ab}(w) \d_a \varphi \d_b \varphi +
    {V}(\varphi) \},\nnn
\ear
in the complex metric background (2.1),
where ${V}(\varphi) \geq 0$ is  a potential,
\be  
{\eta^{ab}}(w) = \diag (w^{-1},1,...,1)
\ee
is the matrix inverse to (2.1) and
\beq  
\eta(w) \equiv   \det (\eta_{ab}(w)) = w, \qquad  w \in \Omega.
\eeq

The action (2.4) is an analytical continuation of the Euclidean
action, (covariantly) defined for $w > 0$, to the domain $\Omega$ (2.2).

From (2.4)-(2.6) we get
\bearr  
{S}(\varphi,w) = \int d^{D} x
\biggl\{ \frac{1}{2} w^{-1/2} (\d_0 \varphi)^2    \nnn
+ \frac{1}{2} w^{1/2} (\vec{\d} \varphi)^2  +
 w^{1/2} {V}(\varphi) \biggr\},
\ear
$w \in \Omega$. The real part of (2.7) satisfies the relations
\bearr  
\re S(\varphi,w) = \cos(\half \arg w)
                     \int d^D x\Bigl\{\half |w|^{-1/2} (\d_0 \varphi)^2 \nnn
\nq + \half |w|^{1/2} (\vec{\d} \varphi)^2 + |w|^{1/2} {V}(\varphi)\Bigr\}
\geq {c_1}(w) {S_{E}}(\varphi),
\ear
where
\be  
S_E(\varphi) = \int d^D x  \Bigl\{ \half
\delta^{ab} \d_a \varphi \d_b \varphi + V(\varphi) \Bigr\}
\ee
is the Euclidean action,
\be  
{c_1}(w) \equiv \cos(\half \arg w) \ \min (|w|^{-1/2}, |w|^{1/2})
\ee
and $\arg w \in (- \pi, \pi)$ is the argument of the complex number $w$.

The action (2.4) generates a chain of $w$-correlators
\bearr  
   \aver{\varphi (x_1)\ldots \varphi(x_n)} (w)
        = \int {\cal D} \mu (\varphi,w)
                             \varphi (x_1) \ldots  \varphi (x_n), \nnn
 \\ \lal
  {\cal D} \mu(\varphi,w)= \frac{{\cal D} \varphi\, \exp(- S(\varphi,w))}
         {\int {\cal D} \varphi\, \exp (-S(\varphi,w))},
\ear
for $n \in \N$ and $w \in \Omega$. In (2.11) and (2.12), the path
integrals are understood in the framework of a certain perturbation scheme
and $R$-operation.

\medskip\noi
{\bf Remark 1.} Due to the relations
\beq  
|\e^{-S(\varphi,w)}| = \e^{-\re S(\varphi,w)}\leq
\e^{-c_1(w) S_E(\varphi)},
\eeq
following from (2.8), one may define for certain  $S(\varphi,w)$
the``complex-valued measure" (charge) (2.12) (at least perturbatively)
for all $w \in \Omega$ if the corresponding Euclidean measure
${\cal D} \varphi \exp(- c {S_{E}}(\varphi))$ exists for any $c > 0$.

\subsection{Free massive scalar field}             

To illustrate the  approach let us consider in detail
the case of a free massive scalar field
(vector, spinor, etc. fields may be certainly considered
as well \cite{I3}), i.e. the potential in (2.4) is
\be        
{V}(\varphi) = \frac{1}{2} m^2 \varphi^2,  \qquad m > 0.
\ee
From (2.7), (2.11), (2.12) and (2.14) we get the  formal
expression for the two-point $w$-correlator
\bearr  
\aver{\varphi(x) \varphi(0)}(w,m) = \int \frac{d^D p}{(2\pi)^D}
            \e^{-ipx}   {D}(p,w,m)        \nnn
\ear
where $w \in \Omega$ and $D(p,w,m)$ is defined in (1.3).
Here and below
\be 
p = (p_a), \quad x = (x^a) \in \R^D, \qquad px = p_a x^a.
\ee
The relation (2.15) should be understood as
\be  
\aver{\varphi(x) \varphi(0)}(w,m) = {\cal F}_p^{-1} [{D}(p,w,m)](x)
\ee
where ${\cal F}_p^{-1} = {\cal F}^{-1}$ is the inverse
Fourier  transformation in ${\cal S}'(\R^D)$, i.e.
in the (dual) Schwarz space of generalized functions (tempered
distributions) \cite{GS,Vl,Io}. Recall that the mapping
${\cal F}: {\cal S}'(\R^D) \to {\cal S}'(\R^D)$
is defined by the relations
\bear  
\aver{{\cal F} T, \varphi} \eql <T, F \varphi>,  \\
          ({F \varphi})(p) \eql \int d^D x\ \e^{ipx}  \varphi(x),
\ear
where $\varphi \in {\cal S}(\R^D)$,
$T \in {\cal S}'(\R^D)$ and
$ F: {\cal S}(\R^D) \to {\cal S}(\R^D)$ is
the Fourier  transformation in the Schwarz space ${\cal S}(\R^D)$.
The space  ${\cal S}(\R^D)$
consists of smooth functions $f: \R^D \to \C $ satisfying the relation
\be  
{p_{n,\alpha}}(\varphi) = {\rm max} \{(1 + x^2_E)^n |D^{\alpha}\varphi|\}
< + \infty
\ee
for all $n \in \N$ and $\alpha = (\alpha_0, \ldots,
\alpha_{D-1}) \in {\bf Z}_{+}^D$. Here ${ \bf Z }_{+} = \{ 0 \} \bigcup
\N $,
$D^{\alpha} = \d_0^{\alpha_0} \ldots
\d_{D-1}^{\alpha_{D -1}}$,
and $\d_i = \d/ \d x^i$.  In (2.20) and below
\be  
x^2_E =  \delta_{ab} x^a x^b , \qquad   p^2_E =  \delta^{ab} p_a p_b.
\ee
The Schwarz space ${\cal S}(\R^D)$ is a locally convex linear
topological space over $\C$ with a topology generated by
the set of seminorms (2.20) \cite{Io}.

Now we prove majorizing inequalities for the $w$-correlator that will play =
a
key role in what follows.

\proclaim Proposition 1. For any  $m > 0$, $w \in \Omega$ and
$p = (p_a) \in \R^D$
\be  
\frac{{c_2^{-1}}(w)}{p^2_E + m^2} \leq |{D}(p,w,m)|
\leq \frac{{c_1^{-1}}(w)}{p^2_E + m^2},
\ee
where ${c_1}(w)$ is defined in (2.10) and
\be  
{c_2}(w) = {\rm max} (|w|^{-1/2}, |w|^{1/2}).
\ee

\medskip\noi
{\bf Proof}. Denote
\bearr  
 X = ({D}(p,w,m))^{-1} = w^{-1/2}p_0^2 {+} w^{1/2}(\vec{p}^{\ 2} + m^2).
 	\nnn
\ear
It is clear that
\be  
|X| \leq {c_2}(w) (p^2_E + m^2).
\ee
We also get:
\bearr  
|X| \geq \re  X \nnn
= \cos(\half \arg w)
[|w|^{-1/2} p_0^2 + |w|^{1/2} (\vec{p}^{\ 2} + m^2)] \nnn
            \inch      \geq {c_1}(w) (p^2_E + m^2).
\ear
The proposition is proved.

\medskip\noi
We denote by  ${{\cal F}_s}(\R^D, \C)$
the vector space of functions $f: \R^D \to \C$,
such that the Lebesgue integral
\be  
\int d^D p\ f(p) (p^2_E + 1)^{-n}
\ee
exists for some $n \in \N$ (in this case the function $f$ is measurable).
The function $f \in {{\cal F}_s}(\R^D, \C)$ generates
the ``slowly increasing" measure ${d \mu}(p) = |f(p)| d^D p$ \cite{Io}.

Let
\be  
I: {{\cal F}_s}(\R^D, \C) \longrightarrow {\cal S}'(\R^D)
\ee
be a canonical embedding defined by the relation
\be  
 \aver {I(f), \varphi} = \int d^D p\ f(p) \varphi(p).
\ee
We call the image of the map (2.28) the subspace of regular
tempered distributions (generalized functions) and denote
\be  
      \im  I = {I}({{\cal F}_s}(\R^D, \C))
                   \equiv {\rm reg} \  {\cal S}'(\R^D).
\ee

\proclaim Proposition 2. For any $w \in \Omega$, $m > 0$,
\be 
D(\cdot,w,m) \in {{\cal F}_s}(\R^D, \C)
\ee
{\rm (see (1.3))} and the corresponding regular distribution
\be 
D(w,m) \equiv {I}(D(\cdot,w,m)) \in {\rm reg}\  {\cal S}'(\R^D)
\ee
is (weakly) holomorphic with respect to $w$ in $\Omega$ (for fixed $m > 0$),
i.e.  $\aver{D(w,m), \varphi}$ is holomorphic with respect to $w$ in
$\Omega$ for any $\varphi \in {\cal S}(\R^D)$ and fixed $m > 0$.

\medskip\noi
{\bf Proof}. For fixed $m > 0$ and  $w \in \Omega$ the function
$D(.,w,m)$  is smooth on $\R^D$ and hence measurable.
The relation (2.31) follows from the right
inequality in (2.22). The holomorphic behaviour of the integral
\beq  
 \aver{ D(w,m), \varphi} = \int d^D p\ D(p,w,m) \varphi(p).
\eeq
follows from the relation
\bearr  
\frac{\d}{\d w} \aver { D(w,m), \varphi } =
        \int d^D p\, \biggl(\frac{\d}{\d w} D(p,w,m)\biggr) \varphi(p)\nnn
\ear
($\varphi \in {\cal S}(\R^D)$) that can be easily verified by a
straightforward calculation using certain uniform estimates in the sectors
\bearr 
\Omega_{r,R,\delta} = \{w: r < |w| < R,\quad |\arg w| <\pi -\delta \},\nnn
\cm   0 < r < R, \cm  0 < \delta < \pi,
\ear
following from  Proposition 1. In this case
\[
  (\d/\d w) D(.,w,m) \in {{\cal F}_s}(\R^D, \C)
\]
and in ${\cal S}'(\R^D)$
\be  
\frac{\d}{\d w} D(w,m) = I\,\Bigl(\frac{\d}{\d w} D(.,w,m)\Bigr)
\ee
for $w \in \Omega$, $m > 0$.

\subsubsection*{Coordinate representation}

The calculation of the inverse Fourier transformation (2.17)
gives us the following expression for the two-point $w$-correlator:
\bearr  
  \aver{\varphi(x) \varphi(0)} (w,m) \nnn
\quad =  (2 \pi)^{-D/2}
      \biggl[\frac{m^2}{x^2(w)}\biggr]^{\nu/2}
 K_{\nu}\Bigl(m \sqrt{[x^2(w)]}\Bigr)
\ear
where $\nu = \nu(D) = D/2- 1$, $m^2 > 0$, $w \in \Omega$ and
\be  
x^2(w) \equiv \eta_{ab}(w) x^a x^b = w (x^0)^2 + \vec{x}^2.
\ee

For any  $w \in \Omega$  and  $m > 0$
\bearr  
\hat{D}(w,m) \equiv {\cal F}^{-1}({D}(w,m))         \nnn
=
{I}(\aver{\varphi(.) \varphi(0)}(w,m))  \in {\rm reg }\ S'(\R^D)
\ear
is a regular distribution, generated by the function of slowly
increasing measure
\be  
\aver{\varphi(.)\varphi(0)}(w,m) \in {\cal F}_{s}(\R^D, \C).
\ee
The inclusion (2.39) follows from the asymptotical relations
\ba 
\nq \aver {\varphi(x) \varphi(0)}(w,m)
    \al \sim \al  A_{D} [x^2(w)]^{-\nu}, \quad D>2,  \\
    \al \sim \al  A_{2} \ln [m^2x^2(w)], \  D =2
\ea
as $x^2_E \to +0$, and
\bearr  
\aver{\varphi(x) \varphi(0)}(w,m) \sim
B_D [x^2(w)]^{a} \exp [-m \sqrt{x^2(w)}], \nnn
\ear
as $x^2_E \to + \infty$, and the inequalities
\be 
{c_1}(w^{-1}) x^2_E \leq |x^2(w)| \leq {c_2}(w^{-1}) x^2_E.
\ee
Here $A_D$, $B_D$ and $a = a_D$ are constants and
${c_i}(w)= {c_i}(w^{-1})$,
$i = 1, 2$ are defined in (2.10) and (2.23).

The relation (2.37) may be obtained by the following three steps.
For $w =1$ (i.e. in the Euclidean case) we use the well-known
Euclidean formula \cite{GJ}
\bearr  
\nq \aver{\varphi(x) \varphi(0) }(w,m)
=\frac{1}{ (2 \pi)^{D/2}}
\biggl(\frac{m^2}{x^2_E}\biggr)^{\nu/2} K_{\nu}\Bigl(m \sqrt{x^2_E}\Bigr).
 \nnn
\ear
Then, performing the dilation $x^0 \mapsto w^{1/2} x^0$,
we get the correlator (2.37) for $w > 0$.
(Note that this dilation generates well-defined endomorphismes of
the linear topological spaces ${S}(\R^D)$ and ${S'}(\R^D)$.)
The distribution $\hat{D}(w,m)$ holomorphically depends on the
parameter $w$ in the domain $\Omega$. It follows from the
holomorphic $w$-dependence of $D(w,m)$ in $\Omega$ and the fact that the
Fourier transformations ${\cal F}$ and ${\cal F}^{-1}$ in  ${S'}(\R^D)$
preserve weak holomorphicity.
The right-hand side of \eq (2.37) defines for $w \in \Omega$
a regular distribution from  ${S'}(\R^D)$ holomorphically
depending on $w$ in $\Omega$.  This implies that the relation (2.37) may be
extended from $ \R_{+}$ to the domain $\Omega$.

\subsubsection*{Proper-time representation}

For the $w$-correlator (1.3) we get the following proper-time
representation ($\alpha$-representation \cite{Bog,Zav,DW}):
\be  
{D}(p,w,m) =  \intl_{0}^{+ \infty} d \alpha\,
                 \e^{ - \alpha w^{1/2} (p^2[w]  +  m^2)},
\ee
for $w \in \Omega$, $m > 0$ and $p \in \R^{D}$. Here
\be  
p^2[w] \equiv \eta^{ab}(w) p_a p_b = w^{-1} (p^0)^2 + (\vec{p})^2.
\ee

Just as in \cite{I3,I4}, we may also consider some general class of
the so-called proper $w$-correlators. A proper $w$-correlator
in the momentum representation has the form
\bearr  
\aver{\Phi_i \Phi_j}(p,w,m) =  P_{ij}(p,w,m) \nnn
  \times
  \intl_{0}^{+ \infty} d \alpha \ f(\alpha w^{1/2})
           \e^{- \alpha w^{1/2} (p^2[w]  +  m^2)}.
\ear
Here $w \in \Omega$, $m > 0$, $p \in \R^{D}$ and $i,j = 1, \ldots, N$
are indices (e. g. vector, spinor, etc). Besides,
\begin{description}
\item[(A)]
    The function $f(\alpha)$ is holomorphic in the domain
         $\{\re  \alpha > 0\}$ and continuous in
         $\{\re  \alpha \geq 0 \} \setminus \{ 0 \}$;
\item[(B)]
    $f(\alpha) = {O}(\alpha^T)$ for $\alpha \to \infty$,
     $\re  \alpha \geq 0$;
\item[(C)]
     there exists $s > -1$ such that, for all $\delta > 0$,
      $f(\alpha) = {O}(\alpha^{s - \delta})$ for $\alpha \to 0$;
\item[(D)]
     all $P_{ij}(p,w,m)$ are polynomials in  momenta $p = (p_i)$ with
     coefficients holomorphically depending on $w$ in $\Omega$
     (and  $P_{ij}(p,w,m)$ are ``$w$-covariant").
\end{description}
The proper $w$-correlators extended to the case $m \geq 0$ form a rather
wide class of $w$-correlators that occur in QFT.

\subsubsection*{The Minkowski space limit, $w \to -1 \pm i 0$}

For the distributions $D(w,m)  \in {\rm reg}\ {\cal S}'(\R^D)$
from (2.32) the limits $w \to -1 \pm i 0$ exist and are
covariant distributions from ${\cal S}'(\R^D)$ defined by the relation
\be 
D(-1 \pm i 0,m) = \mp \frac{i}{2m^2} [2 - \hat{A} L_{\pm}(m)],
\ee
where
\be 
\hat{A} = p_a \d/\d p_a,
\ee
is a continuous operator in  ${\cal S}'(\R^D)$ and
\be 
L_{\pm}(m) \in  {\rm reg}\ {\cal S}'(\R^D)
\ee
are regular distributions generated by the functions
from ${{\cal F}_s}(\R^D, \C)$
\ba 
L_{\pm}(p,m) \eql \ln |p^2_M + m^2| \mp i \pi {\theta}(-p^2_M - m^2)
                                                              \nn
             \eql \ln [ (-1 \mp i0) p^2_0 + \vec{p}^{\ 2} + m^2 ] \nn
             \eql \ln [ p^2_M + m^2 \mp i0 ],
\ea
$p^2_M + m^2 \neq 0$.  Here
\be  
p^2_M \equiv \eta^{ab}(1) p_a p_b = - p^2_0 + \vec{p}^{\ 2}.
\ee
(For $p^2_M + m^2 = 0$ we put $L_{\pm}(p,m) = 0$.)

The relations (2.49) follow from the identity
\bearr  
\nq D(p,w,m) = \frac{w^{-1/2}}{2m^2} \biggl[2 -
p_a \frac{\d}{\d p_a} \ln ( \eta^{ab}(w) p_a p_b {+}m^2 ) \biggr], \nnn
\ear
$w \in \Omega$, ($m > 0$) $p \in \R^{D}$.

Consider the function from ${{\cal F}_s}(\R^D, \C)$
\be  
{D^{\rm cov}_{\pm}}(p, \varepsilon, m) =
\frac{\mp i}{p^2_M + m^2 \mp i\varepsilon},
\ee
$\varepsilon > 0$, $m > 0$. Using the representation
\bearr  
{D^{cov}_{\pm}}(p, \varepsilon, m) \nnn
=\frac{ \mp i}{2(m^2 \mp i\varepsilon)} \biggl[2 -
      p_a \frac{\p}{\p p_a} \ln ( p^2_M  +  m^2  \mp i\varepsilon) \biggr]
\ear
($\varepsilon > 0$), we see that the
covariant regular distributions ${D^{\rm cov}_{\pm}}(\varepsilon, m)
\in {\rm reg}\ {\cal S}'(\R^D)$
corresponding to (2.56) have limits
as $\varepsilon \to + 0$, coinciding with (2.49):
\be 
{D^{\rm cov}_{\pm}}(+ 0, m) =  D(-1 \pm i 0,m).
\ee

\medskip\noi
{\bf Remark 2.} Performing differentiation in (2.49)
and using the well-known relations  in ${\cal S}'(\R)$
(see e.g. \cite{Vl})
\[
(\ln |x|)' = {\cal P} \frac{1}{x}, \quad
x {\cal P} \frac{1}{x} = 1,\quad \theta'(x) = \delta(x) = \delta(-x),
\]
we obtain:
\bearr 
D(p, -1 \pm i 0,m) =
\frac{ \mp i}{p^2_M + m^2 \mp i 0}    \nnn
\qquad = \mp i
\biggl[ {\cal P} \frac{1}{p^2_M + m^2}
\pm i \pi \delta(p^2_M + m^2)\biggr],
\ear
that agrees with the well-known Sokhotski relation.

In the coordinate representation the free
massive scalar field propagator has the following form:
\bearr 
\aver{0|T_{+}({\varphi}(x) {\varphi}(0))|0}(m)=
{{\cal F}_p^{-1}} \biggl[ \frac{- i}{p^2_M + m^2 - i0}\biggr](x). \nnn
\ear
Analogously,
\bearr 
\aver{0|T_{-}({\varphi}(x) {\varphi}(0))|0}(m)
= {{\cal F}_p^{-1}} \biggl[ \frac{i}{p^2_M + m^2 + i0}\biggr](x).    \nnn
\ear
Here $T_{+}(\ldots)$ and $T_{-}(\ldots)$ are chronologically
and antichronologically ordered operator products, respectively
(${\cal F}_p^{-1} = {\cal F}^{-1}$ is the inverse Fourier
transformation in ${\cal S}'(\R^D)$).
>From (2.17), (2.57), (2.59), (2.60) we get:
\bearr 
\aver{0|T_{\pm}(\varphi(x) \varphi(0))|0}(m)
= \aver{\varphi(x) \varphi(0) }( -1 \pm i0, m).  \nnn
\ear
Thus the limits $w = -1 \pm i0$ in 2-point $w$-correlators
correspond to propagators in Minkowski space with the ``right"
and  ``wrong" time directions.

In Ref.\,\cite{I4} we discussed a Feynmann integral corresponding to an
arbitrary connected diagram  (graph) for the theory with the proper
$w$-correlator (2.48) (with $w = - \e^{-i \varepsilon}$).

Analogues of the Bogoliubov-Parasiuk-Hepp-Zim\-mer\-mann theorems were
proved, namely:
\begin{description}
\item[(i)]
the Feynmann integral corresponding to an arbitrary connected
diagram  renormalized in the $\alpha$-representation
exists (as a Lebesgue integral) for all $0 < \varepsilon < 2 \pi$ and
\item[(ii)]
the corresponding   generalized function (distribution) of
external momenta has a limit as $\varepsilon \to +0$ in the appropriate
Schwarz space.  This limit is a covariant distribution.
\end{description}

\subsubsection*{Covariant regularization}

For comparison let us consider the covariant regularization of the propagator
(2.55) ${D^{\rm cov}_{+}}(p, \varepsilon, m)$. This regularization has some
disadvantages as compared with the complex metric regularization (1.3).
First, the covariant regularization does not guarantee the existence of
Feynmann integrals (understood as Lebesgue integrals) even if the
corresponding Euclidean integrals do exist.  The simplest example is: $D=
3$,
\be 
\int \frac{d^3 p}{[(p+q)^2_M + m^2 - i \varepsilon]
[ p^2_M + m^2 - i \varepsilon]}.
\ee
The Lebesgue integral (2.62) does not exist whatever be $q \in \R^3$,
$\varepsilon > 0$, $m > 0$ \cite{I3}. The corresponding Euclidean integral
\be 
\int \frac{d^3 p}{[(p+q)^2_E + m^2][p^2_E + m^2]}
\ee
exists (as a Lebesgue integral)  for all $q \in \R^3$ ($m > 0$).

The covariant regularization (2.55) does not guarantee the
propagator regularity in the coordinate representation.
Indeed, performing the Fourier transformation in (2.59)--(2.60), we obtain:
\bearr  
{\cal F}_p^{-1}\biggl [ \frac{\mp i}{p^2_M + m^2 \mp i \varepsilon}\biggr](=
x)
\nnn
\nq = \frac{1}{(2\pi)^{D/2}}
\biggl(\frac{m^2 \mp i \varepsilon}{x^2_M \pm i0} \biggr)^{\nu/2}
K_{\nu}\Bigl(\sqrt{(m^2 \mp i \varepsilon) (x^2_M \pm i0)}\Bigr)\nnn
\ear
where $\nu = \nu(D) = D/2 - 1$. The relation (2.64) may be obtained fro=
m
(2.37) by performing the replacement $m^2 \mapsto m^2 \mp i \varepsilon$ an=
d
the limit
$w \to -1 \pm i0$. Here we used that fact that the limit
(2.64) in ${\cal S}'(\R^D)$ is unchanged
if the replacement $(-1 \pm i0) (x^0)^2 + \vec{x}^2 \mapsto
x^2_M \pm i0$ is performed. The ``most singular part"  in (2.64) is
\be  
{\rm const} \ (x^2_M \pm i0)^{- \nu}.
\ee
For $D \geq 4$ we have $\nu \geq 1$ and the distribution
(2.65) is singular. Hence (2.64) is singular too.

In \Ref{I1} the complex metric regularization
with $w = -1 \pm i \epsilon$, $\epsilon > 0$ was considered
in the context of the 3-dimensional $\sigma$-model from \cite{Ar}
($n$-field)  and convergence theorems for $\epsilon >0$
were proved. The regularization of propagators with
$w = -1 \pm i \epsilon$ is closely related to
regularization method for pseudo-Euclidean singularities suggested
by W. Zimmermann in \Ref{Z} : $p^2_M \to p^2_M -\epsilon \vec{p}^{\ 2}$.

\section{Multidimensional cosmology with $n$ Einstein spaces}

\subsection{Gravitational model}

Consider the manifold
\beq                          
M = M_{0}  \times M_{1} \times \ldots \times M_{n},
\eeq
with the metric
\bear     
g \eql \e^{2\gamma(x)} \hat{g}^{(0)} +
     \sum_{i=1}^{n} \e^{2\phi^i(x)} \hat{g}^{(i)},  \\
     g^{(0)} \eql g^{(0)}_{\mu \nu}(x)\ dx^{\mu} \otimes dx^{\nu}
\ear
is a metric on the manifold $M_{0}$, and $g^{(i)}$ is a metric on the
manifold $M_{i}$  satisfying
\beq 
{R_{m_{i}n_{i}}}[g^{(i)}] = \lambda_{i} g^{(i)}_{m_{i}n_{i}},
\eeq
$m_{i},n_{i}=1,\ldots, N_{i}$;
$\lambda_{i}= \const$,  $i=1,\ldots,n$.  Thus
$(M_i, g^{(i)})$  are Einstein spaces. In (3.2) we denote
by $\hat{g}^{(\alpha)} =  p_{\alpha}^{*} g^{(\alpha)}$ the
pullback of the metric $g^{(\alpha)}$  to the manifold  $M$
by the canonical projection: $p_{\alpha} : M \rightarrow M_{\alpha}$,
$\alpha = 0, \ldots, n$.  The functions $\gamma, \phi^{i} : M_0 \to\R$
are smooth, $i=1,\ldots,n$.

Consider the gravitational action
\beq           
S  =  S[g] = \frac{1}{2}
\intl_{M} d^{D}x \sqrt{|g|} \{ {R}[g] -  2 \Lambda \}   + S_{\rm GH}
\eeq
where $|g| = \det (g..)$, $S_{\rm GH}$ is the standard Gibbons-Hawking
boundary term \cite{GH} and $\Lambda$ is the cosmological constant.

The field equations for the action (3.5) (Einstein equations)
\be 
R_{MN}[g] - \half g_{MN} R[g]  =  - \Lambda g_{MN}
\ee
are (for $D\ne 2$) equivalent to
\be 
R_{MN}[g]  =  2 \Lambda g_{MN}/(D-2),
\ee
where  $D = \sum_{k=0}^{n} N_k = \dim M$
is the dimension of the manifold (3.1), $N_k = \dim M_k$, $k=0,\ldots, n$.

Although the cosmological case $N_0 = 1$ will be our main subject,
we shall need the more general non-exceptional case $N_0 \neq 2$.
In this case we put, just as in \cite{Ber,IM6},
\be 
\gamma = {\gamma}_{0}(\phi) =
\frac{1}{2-N_0}  \sum_{i =1}^{n} N_i \phi^i.
\ee

It may be shown (see \cite{Ber,RZ,IM6}) that \eqs
(3.6), (3.7)  for the metric (3.2) with $\gamma$ from (3.8) are equivalent
to the  equations of motion for the $\sigma$-model
\bearr                                                  
S_0[g^{(0)},\phi] = \frac{1}{2}
\intl_{M_0} d^{N_0}x \sqrt{|g^{(0)}|} \{ {R}[g^{(0)}] \nnn
\cm - G_{ij} g^{(0) \mu \nu}
\d_{\mu} \phi^i  \d_{\nu} \phi^j -  2 V(\phi) \}
\ea
where $|g^{(0)}| = \det (g^{(0)}..)$,
\be 
G_{ij} = N_i \delta_{ij} + \frac{N_i N_j}{N_0 -2}
\ee
are components of the ``midisuperspace" (or target space) metric on  $\R^{n
}$
\be 
G = G_{ij} d \phi^{i} \otimes d \phi^{j}
\ee
and
\bearr 
V(\phi) = \Lambda \e^{2 \gamma_0(\phi)}-
\frac{1}{2}
\sum_{i =1}^{n} \lambda_i N_i \e^{-2 \phi^i + 2 \gamma_0(\phi)}
\ear
is the potential. (In \cite{RZ} authors start with $\gamma =0$.)

We note that \cite{IMZ}
\be 
\det (G_{ij}) = N_1 \ldots N_n \frac{2 - D}{2 - N_0} \neq 0.
\ee

\subsection{Multidimensional cosmology}

Now consider the cosmological case
\be 
N_0 =1, \qquad  g^{(0)} = - dt \otimes dt.
\ee
In this case (3.8) corresponds to the harmonic-time gauge
and the minisuperspace metric
\be 
\hat{G}_{ij} = N_i \delta_{ij} - N_i N_j
\ee
has the pseudo-Euclidean signature $(-,+, \ldots ,+)$ \cite{IM1,IMZ}.
The equations of motion for the cosmological model under consideration
are equivalent to the Lagrange equations for the Lagrangian
\begin{equation} 
L = \frac{1}{2} \hat{G}_{ij} \dot{\phi}^{i} \dot{\phi}^{j} - V
\end{equation}
with the energy constraint \cite{IMZ}
\beq 
E = \Half \hat{G}_{ij} \dot{\phi}^{i} \dot{\phi}^{j} +  V = 0.
\eeq
(For exact solutions of the Einstein equations see e.g.
\cite{IM1}--\cite{GIM2}.)

\subsection{The quantum case}

The quantization of the zero-energy constraint (3.17) leads to the
Wheeler-DeWitt (WDW) equation in the harmonic time gauge \cite{IMZ}
(see also \cite{IM5,BIMZ})
\begin{equation} 
 \hat{H} \Psi \equiv \biggl[ - \frac{1}{2} \hat{G}^{ij}
\frac{\d}{\d \phi^{i}} \frac{\d}{\d \phi^{j}}+ V \biggr] \Psi = 0,
\end{equation}
where
\be 
\hat{G}^{ij} = \frac{\delta_{ij}}{N_i} + \frac{1}{2 - D}
\ee
are components of the matrix inverse to $(\hat{G}_{ij})$ in (3.15).

\medskip\noi
{\bf Third quantized model.} The WDW equation (3.18) corresponds to the
action of so-called "third-quantized" cosmology \cite{Rub}-\cite{GidS}
(see also \cite{Pel1}-\cite{Zh5}, \cite{IM5})
\be 
S = \frac{1}{2} \int d^{n+1} z\ \Psi \hat H \Psi.
\ee
We may quantize (3.20) and study the processes of ``creation" and
``interaction" of ``multidimensional universes" \cite{Zh1,Zh5}.

\medskip\noi{\bf Diagonalization}. The operator $\hat H$ (3.18) may be
diagonalized by the linear transformation
\be 
\varphi^a = S^a_i \phi^i,
\ee
where
\be 
S^a_i  \delta_{ab}  S^b_j = G_{ij},
\ee
$a, b = 0, \ldots, n-1$; $i, j = 1, \ldots, n$. In this case the
midisupermetric  (3.11) reads:
\be 
     G = \delta_{ab} d \varphi^{a} \otimes d \varphi^{b} .
\ee

An example of such a diagonalization is \cite{IM1,IMZ}
\bear        
\varphi^0 \eql  q^{-1} \sum_{i=1}^{n} N_i \phi^i, \\
\varphi^{\hat{b}} \eql
\Big[N_{\hat{b}-1}/(\Sigma_{\hat{b}-1} \Sigma_{\hat{b}})  \Big]^{1/2}
\sum_{j= \hat{b}}^{n} N_j (\phi^j {-} \phi^{ \hat{b}-1})   ,
\ear
$\hat{b} = 1, \ldots,n - 1$, where
\be  
q \equiv \biggl[\frac{(D - 1)}{(D -2)} \biggr]^{1/2},\qquad
\ \Sigma_a =  \sum_{j=a}^{n} N_j.
\ee

Here we consider a more general class of diagonalizations (3.21)
satisfying (3.24) or, equivalently,
\beq  
S^0_i = q^{-1} N_i,      \cm  i = 1, \ldots, n.
\eeq
One can show using the relations from \cite{IM6}
that the potential (3.12) is written in the new variables as
\be                           
V = \Lambda \e^{2q \varphi^0} +
     \e^{2q^{-1} \varphi^0} V_* (\vec{\varphi}_{*})
\ee
where
\be 
{V_{*}}(\vec{\varphi}_{*}) =\sum_{i =1}^{n} ( - \half \lambda_{i} N_i)
\exp(\vec{u}^{(i)}_{*} \vec{\varphi}_{*}),
\ee
and the vectors  $\vec{u}^{(i)}_{*} \in \R^{n-1}$ satisfy the relations
\be 
\vec{u}^{\ (i)}_{*} \vec{u}^{\ (j)}_{*} =
4 \biggl( \frac{\delta_{ij}}{N_i} + \frac{1}{1 - D} \biggr).
\ee
The operator  $\hat{H}$ in the new variables reads
\be 
\hat{H} = - \eta^{ab} \frac{\d}{\d \varphi^{a}} \frac{\d}{\d \varphi^{b}}
+ V(\varphi),
\ee
where $V = V(\varphi)$ is defined in (3.28) and
\be 
\eta^{ab} = \eta_{ab} =
{\eta^{ab}}(-1) = \diag (-1, 1, \ldots, 1).
\ee

\medskip\noi
{\bf Complex dimensions.} Consider the simplest case $V = 0$. Then
the third quantized cosmological model is equivalent
to the theory of a free massive field in $\R^{n-1}$ with
the Minkowski metric (3.32). There are at least two
possibilities of complex metric regularization for (3.15).

First, we may consider the gravitational model (3.1)-(3.12)
with the time manifold $M_0$ of dimension $N_0 > 2$
(when the midisuperspace metric is Euclidean),
then perform the analytical continuation to
the region $N_0 < 2$ and consider the limit
\be 
                  N_0 \to 1 - i0.
\ee
(see (3.13).)

The second possibility \cite{IM2,IM5} is as follows:
we put $N_0 = 1$ and consider (formally) the range of ``small"
dimensions $N_i$
\be 
N_i >0, \qquad \sum_{i =1}^{n} N_i <1,
\ee
where the minisupermetric (3.15) is Euclidean, and then perform the
analytical continuation to the original $N_i$  considering the limits
\beq     
N_i  - i0,  \cm  i = 1, \ldots, n.
\eeq
In this context it is worth noting that there already exist
studies of multidimensional models with dimension ``dynamics"
\cite{BMR}. In this model the minisupermetric signature may change
(if ``small" dimensions are considered).

\section{Concluding remarks}

In the cosmological model under study the minisupermetric
does not depend on the signatures of the manifolds $(M_i,g^i)$.
There exist models where such a dependence takes place.
In \cite{IM7} a gravitational model with several forms
and dilatonic fields was considered. It generalizes the
``pure gravitational" model \cite{IM6} of Subsec.\,3.1.
It turns out that the part of the midisuperspace metric corresponding
to the forms crucially depends on the signatures of the manifolds
$(M_i, g^i)$,  $i =1, \ldots, n$. It will of interest to
consider the Wick rotaion in this model \cite{IM7}.

Another problem of interest is connected with some cosmological models with
spinors (e.g. supersymmetric models) were the Wick rotation (both in space
and in midisuperspace) may be non-trivial  \cite{N}.

\Acknow
{This work was supported in part by the Russian State Committee for
Science and Technology, Russian Fund for Basic Research
(project N 95-02-05785-a)  and by DFG grant 436 RUS 113/7.
The author thanks K.A.Bronnikov for valuable comments. The author is also
grateful to M. Rainer for hospitality during author's  stay in Potsdam
University, reading the text and valuable comments.}

\end{document}